\documentclass[prl,twocolumn,nofootinbib,aps,preprintnumbers,amsmath,amssymb,superscriptaddress,floatfix]{revtex4-1}
\usepackage{graphicx}
\usepackage{color}
\usepackage{bm}
\usepackage{amsmath}
\usepackage{xspace}
\usepackage[normalem]{ulem}
\usepackage{units}
\usepackage{dcolumn}
\usepackage{paralist}
\usepackage{natmove}
\usepackage{natbib}
\usepackage{hyperref}

\begin{document}

\title{Anisotropic In-Plane Phonon Transport in Ultrathin Silicon Membranes Guided by Nano-Surface-Resonators}
\author{Sanghamitra~Neogi}
\email{sanghamitra.neogi@colorado.edu}
\affiliation{Ann and H.J. Smead Aerospace Engineering Sciences, University of Colorado Boulder, Boulder, CO 80303, USA}

\author{Davide~Donadio}
\affiliation{Department of Chemistry, University of California Davis, One Shields Ave. Davis, CA, 95616}

\date{\today}

\begin{abstract}

Anisotropic phonon transport along different lattice directions of two-dimensional (2D) materials has been observed, however, the effect decreases with increasing the thickness beyond a few atomic layers. Here we establish a novel mechanism to induce anisotropic phonon transport in quasi-2D materials with isotropic symmetry. The phonon propagation is guided by resonance hybridization with surface nanostructures. We demonstrate that the thermal conductivity of 3 nm-thick silicon membrane with surface nanofins is greater by $\sim50\%$ parallel to the fins than that perpendicular to the fins. 
\end{abstract}

\maketitle

\section{introduction}
Heat conduction in bulk materials is well described by Fourier's law~\cite{baron1822theorie} in terms of the proportionality between the heat current and the local temperature gradient. The coefficient of proportionality defines the thermal conductivity, $\kappa$: an intensive property, independent of materials dimension and geometry. However, recent works heightened the debate over the applicability of Fourier's law to describe heat conduction in low-dimensional materials~\cite{lepri2003thermal,dhar2008heat}. 
In fact the thermal conductivity (TC) at the nanoscale can differ significantly from the bulk counterparts~\cite{liu2012anomalous,ghosh2010dimensional,balandin2011thermal,xu2014length, chang2008breakdown,meier2014length,hsiao2013observation,yang2010violation}, with major consequences for the application of low-dimensional materials, such as graphene~\cite{ghosh2008extremely,ghosh2010dimensional,balandin2008superior,balandin2011thermal,xu2014length}, transition metal dichalcogenides~\cite{wang2012electronics,xu2013graphene,mak2010atomically},  nanotubes~\cite{chang2008breakdown}, and nanowires~\cite{hsiao2013observation,yang2010violation}, in electronic, optoelectronic, and phononic/thermal devices~\cite{pop2010energy,fukuda1991reliability,yan2012graphene,neto2009electronic,wang2012electronics,bertolazzi2013nonvolatile,georgiou2013vertical,li2014black,Sood:2018ci}. 

Anomalous length dependent TCs have been observed in one-~\cite{chang2008breakdown,Hsiao:2013gz} and two-dimensional nanomaterials~\cite{xu2014length}, beyond 10 $\mu$m and even up to millimeters~\cite{Lee:2017he}. Strong anisotropy of TC is known to exist in van der Waals layered nanomaterials when comparing the in-plane and cross-plane directions~\cite{slack1962anisotropic,balandin2011thermal}. Anisotropic in-plane TCs have recently been observed in few-layer black phosphorus~\cite{luo2015anisotropic,lee2015anisotropic}, and theoretically predicted for phosphorene~\cite{jain2015strongly}, borophane (hydrogenated boron sheet)~\cite{liu2017anisotropic}, arsenene~\cite{Zeraati:2016in}, and silicene~\cite{zhou2018anisotropic}. Anisotropy is also predicted to develop along different lattice directions in the basal plane of graphene and MoS$_2$~\cite{xu2009intrinsic,aksamija2011lattice,liu2013phonon,Chen:2019dg}. This high anisotropy is attributed to the in-plane structural asymmetry that reflects into the direction-dependent phonon dispersion, group velocity, and phonon-phonon scattering. Surface reconstructions are shown to create in-plane TC anisotropy in 2-4 atomic layer thick silicene, and the effect monotonically decreases with increasing thickness~\cite{zhou2018anisotropic}. To date, however, in-plane anisotropic thermal conductivity has not been reported for thin films of materials with isotropic symmetry (cubic or hexagonal).  

In this Letter, we introduce a new concept to engender anisotropic in-plane phonon transport in thin films or membranes of isotropic materials, by hybridizing the membrane phonons with resonances introduced by surface nanostructures. 
Using atomistic lattice dynamics and classical molecular dynamics simulations we illustrate that phonon propagation in suspended silicon membranes ($>20$ atomic layers) with surface ``fins" is anisotropic and guided by the fin geometry. We demonstrate the concept on silicon membranes with periodic surface nanoscale pillars and fins. Silicon is chosen because of its wide use in a broad range of technological applications and ease of fabrication, however the concept is generically applicable to other materials.
The effect of dimensional reduction~\cite{hochbaum2008enhanced,donadio2009atomistic,lim2012quantifying,xiong2016blocking,davis2014nanophononic,neogi2015thermal,neogi2015tuning,xiong2017native,honarvar2016thermal,honarvar2018two} and surface roughness, due to oxidation~\cite{zushi2015effect,neogi2015tuning,xiong2017native}, amorphization~\cite{hochbaum2008enhanced,donadio2009atomistic,lim2012quantifying} or fabricated nanostructures~\cite{davis2014nanophononic,neogi2015thermal,xiong2016blocking,honarvar2016thermal,honarvar2018two} are extensively discussed to reduce $\kappa$ of low-dimensional silicon, yielding potential thermoelectric applications~\cite{mangold2016optimal,Donadio:201935}. In particular, nanopillar resonators on silicon thin-films  exhibit unique subwavelength phonon properties at the nanoscale, resulting in strong $\kappa$ reduction~\cite{davis2014nanophononic,honarvar2018two}.  
Here we unveil a new mechanism to tune phonon propagation direction and localization, and, as a result, directionality of heat transport in thin films, using nano-engineered surfaces. Through a systematic analysis of phonon dispersion, group velocity, localization, and lifetime, we highlight the role of resonance hybridizations to induce anisotropic phonon transport in the nanostructured membranes.

\begin{figure}
\begin{center}
\includegraphics[width=\linewidth]{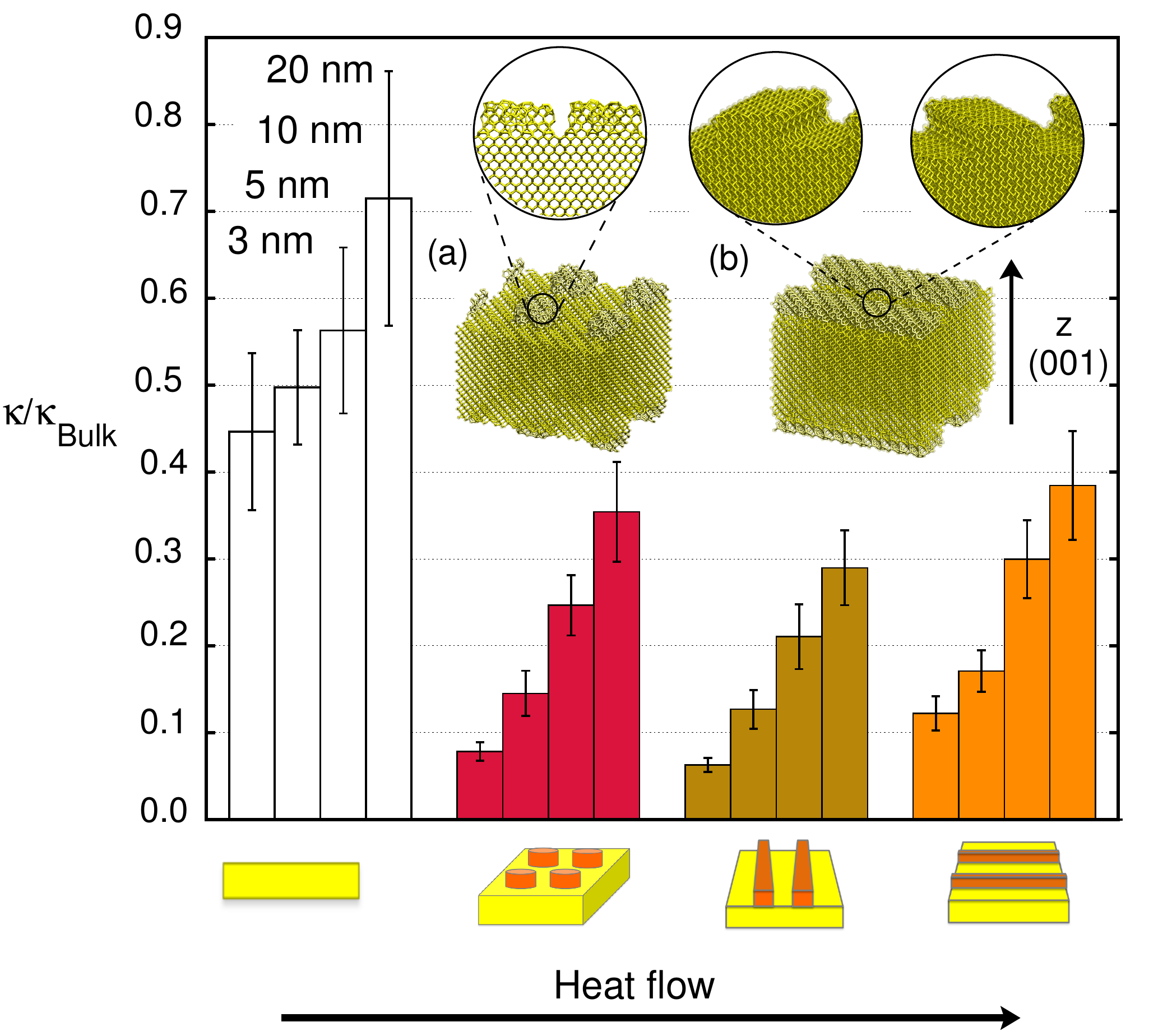}
\caption{{\bf Ratio between the room-temperature thermal conductivity of bulk silicon and silicon membranes ($\kappa/\kappa_{\text{Bulk}}$) with atomistic smooth surfaces (white), surface nanoscale pillars (red) and fins (gold, orange)} with thicknesses 3 nm, 5 nm, 10 nm and 20 nm, respectively. Cartoons in the bottom refer to the corresponding configurations above them. The two rightmost sets depict in-plane $\kappa$ tensor components in the direction shown by the arrow at the bottom of the figure. The results illustrate the anisotropy in the in-plane $\kappa$ as a function of the fin orientation. (Inset) Representative microscopic configurations of silicon membranes with nanoscale silicon (a) pillars, (b) fins at the top and bottom surfaces.}
\label{fig:kappa}
\end{center}
\end{figure}

We show the representative atomistic model configurations of suspended silicon membranes with surface nanoscale (a) pillars and (b) fins in Fig.~\ref{fig:kappa} (inset). We prepare the membrane configurations by cleaving bulk Si supercells along the [001] direction to construct pristine surfaces as well as with monolithic pillars and fins at both surfaces. The supercells are constructed by replicating a Si cubic conventional cell (CC): the pristine membrane supercells consist of $8\times8\times n_z$ CCs, the surface nanopillars $4\times4\times2$ CCs and nanofins $4\times8\times2$ CCs, respectively, where $n_z$ is the number of unit cells in the $z$-direction. We investigate single-crystalline membranes with four different thicknesses: 3 nm, 5 nm, 10 nm and 20 nm, with $n_z = 6, 10, 20$ and 37, respectively. Both the pristine and structured membranes are $2\times1$ surface reconstructed forming rows of dimers, to minimize the number of dangling bonds~\cite{appelbaum1976si}. The supercells were heated to 1500 K for 2 ns and then quenched to 300 K by Langevin dynamics with a cooling rate of $\sim 10^{11}$ K/s to obtain equilibrated configurations. The surface nanopatterns relax into periodic features with spacing $\sim 2$ nm, height $\sim 1$ nm and area $2\times2$ nm$^2$ (pillars) or $2 \times $(sample length) nm$^2$ (fins), respectively. The chosen model surface nanostructures mimic the rough surfaces of fabricated silicon wafers closely~\cite{tong1997role,feenstra1985surface}.

In order to gain microscopic understanding of heat transport in the nanostructured membranes, we performed a series of equilibrium molecular dynamics (EMD) simulations in which the interatomic interactions are modeled using the empirical Tersoff potential~\cite{tersoff1989modeling} (see details~\cite{mddetails}). Tersoff potential reproduces to a fair extent the TC of both crystalline and amorphous silicon~\cite{he2012lattice,Isaeva:2019dm}. The EMD simulations were carried out using LAMMPS~\cite{plimpton1995fast}. The TC is computed from the fluctuations of the heat current in the EMD simulations, using the Green-Kubo relation~\cite{zwanzig1965time,Schelling:2002jl}. We verified that a $x-y$ periodic cell dimension of $16\times16\times n{_z}$ CCs, i.e. $(8.7 \times 8.7 \times\ \text{thickness})$ nm$^3$, is sufficient to achieve well converged values of $\kappa$, by comparing results with cells as large as $64\times64\times n{_z}$ CCs $\sim(34.8 \times 34.8 \times \text{thickness})$ nm$^3$~\cite{neogi2015thermal}. 
We then performed a spectral analysis of the heat carrying vibrations of the membranes and computed modal contributions to $\kappa$, implementing the Boltzmann transport equation (BTE) approach in the single mode relaxation time approximation (RTA). We adopted a perturbative approach to compute the phonon-phonon interactions, and neglect contributions from four or higher-order scattering processes~\cite{srivastava1990physics}. The supercells used in BTE approach are pristine membranes with $8\times8\times6$ CCs containing 3072 atoms, and membranes with 1 nm-high pillars ($4\times4\times2$ CCs) and fins ($4\times8\times2$ CCs) containing 3584 and 4160 atoms, respectively. All calculations refer to systems at room temperature (300 K).

\begin{figure*}
\begin{center}
\includegraphics[width=1.0\linewidth]{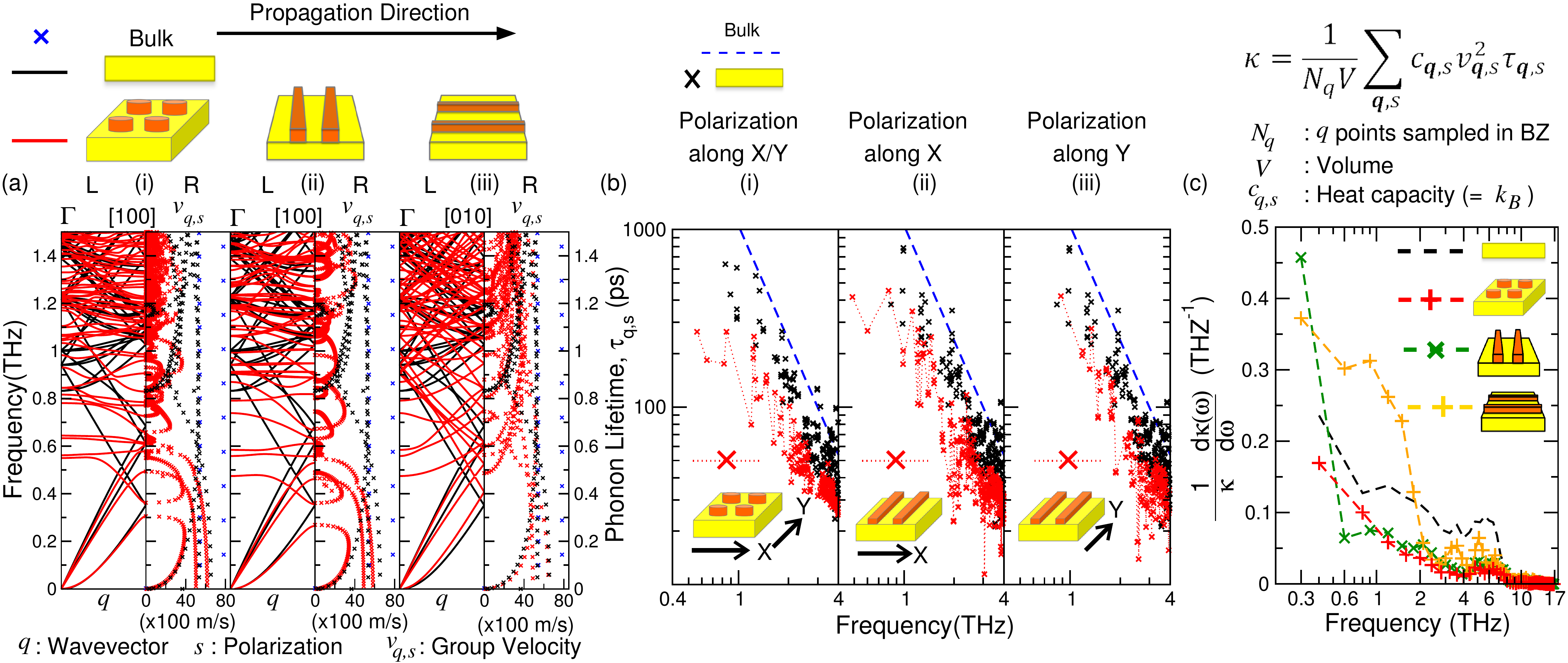}
\caption{{\bf Spectral analysis of phonon propagation in nanostructured 3nm-thick silicon membranes}: (a) {\bf Phonon dispersion and group velocities} in membranes with periodic nanoscale surface (i) pillars and (ii, iii) fins, calculated within the harmonic approximation along different symmetry directions: (L) dispersion and (R) group velocity. The corresponding geometries are depicted with cartoons on top of the figure. The phonon properties are shown along (ii) perpendicular and (iii) parallel to the fins in the membranes with nanofins. The black and red lines in the L panels correspond to membranes with pristine and nanostructured surfaces, respectively. The flattening of modes due to resonances can be clearly discerned in the two panels ((i), (ii)) while ``guide"-like modes are visible in panel (iii). The blue, black and red X's in R panels represent bulk Si, and membranes with pristine and nanopatterned surfaces, respectively. The R panels reflect the direct effect of phonon hybridization on group velocities. (b) {\bf Phonon lifetimes} in membranes with (i) pillars and (ii, iii) fins, calculated with anharmonic lattice dynamics. Blue dashed lines indicate the fitting of $\tau$ of bulk silicon to ~$1/\omega^{2}$. (c) {\bf Differential modal contribution to thermal conductivities} of nanostructured membranes, calculated with Boltzmann transport equation, Within the single mode relaxation time approximation, shown on top of the figure. 
The signatures of anisotropic propagation properties are present across all the panels.}
\label{fig:dispersion_nanostructures}
\end{center}
\end{figure*}
Figure~\ref{fig:kappa} summarizes the in-plane $\kappa$ of nanopatterned membranes computed with the Green-Kubo relation. The $\kappa$ values are scaled with respect to the bulk Si reference: $\kappa_{\text{Bulk}} = 197\pm20$ W/m-k~\cite{he2012lattice}. The presence of surface nanostructures result in reduced TCs (red, gold, and orange blocks) compared to pristine membranes (white blocks) of similar thicknesses, confirming previous works~\cite{davis2014nanophononic, honarvar2016thermal, honarvar2016spectral, honarvar2018two, neogi2015thermal,neogi2015tuning, xiong2017native}. Resonant modes of nanopillars hybridize with the underlying phonon dispersions of the base Si membranes and such couplings drastically lower the in-plane TC~\cite{davis2014nanophononic}. 
With a fixed nanostructure configuration, $\kappa$ reduction increases with decreasing the membrane thickness, reaching a maximum 13-fold and 16-fold reduction for the 3 nm-thick membranes with nanopillars and fins, respectively. 
This is due to the increase of the surface-to-volume ratio and consequentially, the mode coupling between the membrane and the resonators~\cite{honarvar2016thermal}. Remarkably, the nanofins break the membrane $x-y$ in-plane symmetry engendering strongly anisotropic TC. The TC parallel to the nanofins ($\kappa_{||}$) is much higher than the  perpendicular direction ($\kappa_\perp$) (see Figure~\ref{fig:kappa}). $\kappa_{||}$ exceeds $\kappa_\perp$ by $\sim$ 50\% in 3 nm and by $\sim$ 25-30\% in the 5 nm, 10 nm and 20 nm-thick membranes.

\begin{figure}
\includegraphics[width=\linewidth]{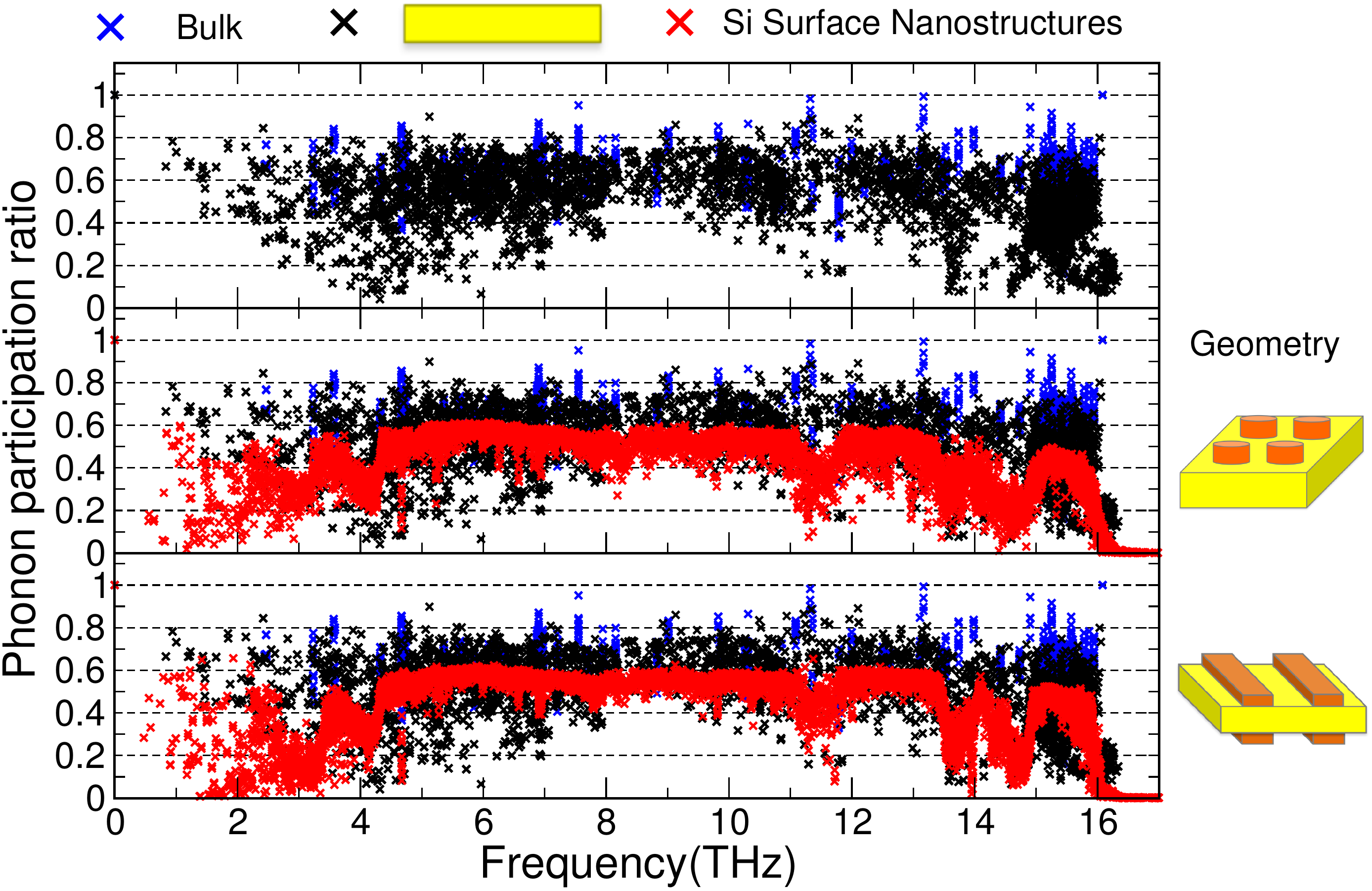}
\caption{Participation ratio ($p_i$) of phonons of nanostructured silicon membranes: surface nanostructures are instrumental in localizing phonon modes, especially in the low frequency region. $p_i$ indicates the fraction of atoms participating in a given eigenmode and is defined as $p_i^{-1}=N\sum_n [\sum_\alpha e^*_{\alpha,n}(i)e_{\alpha,n}(i)]^2$, where $e_{\alpha,n}(i)$ is the $\alpha$ component of the mode $i$ relative to the coordinate of the atom $n$.}
\label{fig:ppratio}
\end{figure}
Hereafter, we aim at unraveling the microscopic origin of the observed anisotropic thermal transport, and its relation to phonon hybridization with surface resonances. 
To this end, we perform a detailed spectral analysis using anharmonic lattice dynamics. Figure~\ref{fig:dispersion_nanostructures}(a) shows the dispersions $(\omega_s({\bf q}))$ and group velocities $({\bf v}_{{\bf q},s})$ of phonons of 3 nm-thick nanostructured Si membranes along different symmetry directions. We chose the 3 nm-thick membrane to analyze the strong impact of mode hybridization given a high surface-to-volume ratio, and to minimize computational cost. 
We highlight the low-frequency phonons since they provide the dominant contribution to TC, thus accounting for the trends exhibited in Fig.~\ref{fig:kappa}. 
Acoustic phonon dispersions at small wavevectors are not affected by surface features Fig.~\ref{fig:dispersion_nanostructures}(a)(i-L, ii-L, iii-L), however, flat locally resonant branches appear in (i-L) and (ii-L). The coupling between surface resonances and membrane phonons leads to flattening of the phonon branches at the resonant frequencies, across the entire spectrum~\cite{davis2014nanophononic}. The lowest frequency of the flat branches can be tuned by optimizing the resonances (e.g., by changing the linear dimensions of the resonators)~\cite{honarvar2016thermal}. Remarkably, the dispersion relations parallel to the fins do not exhibit flat modes, but rather presence of ``guide"-like modes (Fig.~\ref{fig:dispersion_nanostructures}(a) (iii-L))~\cite{koshelev2016interplay}. 

The mode flattening has a direct influence on the phonon group velocities (${\bf v}_{{\bf q},s} = d\omega_s/d\mathbf{q}$), that are significantly smaller in nanostructured membranes than those in the pristine membrane (Fig.~\ref{fig:dispersion_nanostructures}(a)R) panels, Figure S1 in the Supplemental Material). Such drastic reduction of ${\bf v}_{{\bf q},s}$ is the main cause for the observed TC reduction in pillared membranes and perpendicular to the fins (along [100]). In both the cases ${\bf v}_{{\bf q},s}$ are most significantly reduced at the frequencies where the local surface resonances couple with the membrane phonon modes (Fig.~\ref{fig:dispersion_nanostructures} (a) (i-R) and (ii-R)). 
In contrast, ${\bf v}_{{\bf q},s}$ parallel to the fins (along [010]), remain significantly larger (iii-R). 
This aspect reveals a direct advantage of the fin geometry to induce anisotropy and offers new promises to guide phonon propagation in quasi-2D materials through the exploitation of surface nanostructure geometry. 

The altered dispersions also impact the phonon relaxation times ($ \tau_{{\bf q},s}$) in nanostructured membranes (Fig.~\ref{fig:dispersion_nanostructures}(b)). $\tau_{{\bf q},s}$ diverges at low frequencies as $\sim1/\omega^\alpha$. For bulk Si (blue), $\alpha\sim 2$~\cite{Klemens_1955}, whereas for pristine membranes (black) $\tau_{{\bf q},s}$ exhibits a slightly weaker divergence, $\alpha\sim 1.5$. 
In comparison, $\tau_{{\bf q},s}$'s are considerably lower in nanopatterned membranes than in pristine membranes at low frequencies ($\leq $ 3 THz) and show some remarkable features. The $\tau_{{\bf q},s}$'s of higher frequency phonon modes are not affected significantly. A similar observation is reported for oxidized silicon membranes that native oxide induces resonance to strongly suppress the phonon mean free paths below 4 THz~\cite{xiong2017native}. Even more remarkable impact of the resonances can be noted when we categorize $\tau_{{\bf q},s}$ according to the phonon polarizations. We identify the polarizations from the maximum projection of the eigenvectors $(e_{\alpha,n}(i))$ of the dynamical matrix along the three Cartesian axes, $\alpha$ (Fig.~\ref{fig:dispersion_nanostructures}(b)). Here $n$ denotes atoms and $i\equiv ({\bf q},s)$. The $\tau_{{\bf q},s}$'s of low frequency phonons, polarized along X or Y directions, show identical behavior in the nanopillared membrane, and reach a plateau ($\leqslant 1$ THz) (Fig.~\ref{fig:dispersion_nanostructures}(b)(i)). However, the low-frequency phonons, polarized parallel to the fins are long-lived while $\tau_{{\bf q},s}$ of those perpendicular to it reach a plateau (Fig.~\ref{fig:dispersion_nanostructures}(b)(ii) and (iii)). This suggests that mode hybridization not only affects ${\bf v}_{{\bf q},s}$ but impacts the second-order phonon-phonon scattering processes as well. 

We combined ${\bf v}_{{\bf q},s}$ and $\tau_{{\bf q},s}$ to compute the modal contribution to $\kappa$. We integrate contributions from all phonons with frequencies between $\omega$ and $\omega + d\omega$ and compute the $\omega$-dependent differential TC accumulation function. We normalize the TC accumulation by the total $\kappa$ to highlight the weight of the modal contributions as a function of frequency, and to facilitate the comparison between different systems of interest (Fig.~\ref{fig:dispersion_nanostructures}(c)). The dimensional reduction minimally reduces the contributions from low frequency phonon modes of the pristine membrane. In comparison, the mode contributions are severely reduced across all frequency ranges in the nanostructured membranes. The $\kappa$ accumulations perpendicular to fins show similar trend as nanopillared membranes, which is consistent with geometric similarity. However, the low frequency modes parallel to the fins provide a greater contribution due to relatively high ${\bf v}_{{\bf q},s}$ and higher $\tau_{{\bf q},s}$ compared to the perpendicular direction, thus corroborating the MD results shown in Fig.~\ref{fig:kappa}. These results establish the fundamental mechanisms to induce anisotropic spectral contributions to TC of nanostructured thin films by surface nanoscale engineering. 

To further characterize the propagation of the hybridized phonon modes, we probe phonon localization in the nanostructured membranes by computing the participation ratios~\cite{allen1999diffusons,feldman1993thermal} (Figure~\ref{fig:ppratio}). Surface reconstruction does not impact the extended nature of the phonons of the pristine membrane (top). However, the resonance hybridizations strongly reduce the participation of atoms in the low-frequency phonon modes ($\omega\leq 2$ THz) resulting in localization and reduced propagation (Fig.~\ref{fig:ppratio} (middle and bottom)). The localizing effect is more prominent in nanopillar-ed configurations than fins, thus leading to lower group velocities and ultimately larger reduction of $\kappa$. Note that the participation ratio is not resolved with respect to polarization and thus does not offer any information regarding the directional phonon propagation. In recent years, remarkable evidences of phonon localization  have been demonstrated in nanostructured materials due to the introduction of multiple scattering and interference mechanisms with periodicity on the order of the length scale of the propagating waves~\cite{luckyanova2018phonon,hu2018randomness,hu2019disorder}. Here we demonstrate a novel mechanism to induce localization via resonances, without introduction of internal scattering mechanisms within the material. The surface structures induce numerous resonances, each of which may hybridize with the host phonons, and thus enable emergence of unique propagating character of phonons.

In this Letter, we establish a novel mechanism to engender directional, anisotropic vibrational mode propagation in quasi-2D materials. The modal propagation is guided by resonance hybridizations induced by surface nanostructures. We demonstrate the mechanism by analyzing phonon properties of ultra thin silicon membranes with periodic nanoscale pillars and fins on surfaces. Flat modes appear in the dispersion across the whole frequency spectrum in nanopillared membranes and membranes with fins, perpendicular to the fins. Consequentially, the group velocities are strongly reduced. Remarkably, ``guide"-like modes appear parallel to the fins, resulting in less reduction in group velocities. Furthermore, the lifetimes of low frequency phonons reach a plateau in the pillar and across-fin directions, while parallel-fin directions show weak divergence. The thermal conductivities computed with BTE-SMRT and EMD approaches reflect the reduction and anisotropy. $\kappa_\parallel$ exceeds $\kappa_\perp$ by $\sim 50\%$ in 3 nm-thick membranes. Our results establish the fundamental mechanisms to localize and guide phonons of quasi-2D materials using surface nanoscale engineering. 

The primary advantage of the configurations is that they will pose minimal challenges in implementation to existing devices and novel materials architectures. The periodic nanopillars or fins could be fabricated using dry etching~\cite{chekurov2009fabrication}, metal assisted chemical (wet) etching~\cite{huang2008extended}, dislocation-driven mechanism~\cite{bierman2008dislocation}, and vapor-liquid-solid processes~\cite{dick2004synthesis,jung2007synthesis}. We anticipate that our results showing direct relationship between engineered nano-surface-resonators and phonons will open up new research directions to control phonons of existing and new technology-enabling nanomaterials for a broad range of applications, including heat dissipation in nanoelectronics.

\section{Acknowledgements}
This work is partly funded by the European Commission FP7-ENERGY-FET project MERGING with contract number 309150. We acknowledge financial support from MPG under the MPRG program and the provision of computational facilities and support by Rechenzentrum Garching of Max Planck Society (MPG).

\bibliography{siLiterature}

\begin{thebibliography}{74}
\bibitem [{mdd()}]{mddetails}%
  \BibitemOpen
  \href@noop {} {}\bibinfo {note} {In the equilibrium MD simulations, the
  equations of motion are integrated with a timestep of 0.5 fs so to guarantee
  energy conservation over simulation times of several tens of ns. The initial
  velocities were set to 300 K and the systems were coupled to a Nos\'e-Hoover
  thermostat for 1 ns to decorrelate the systems from the initial
  configurations. The thermostat was then decoupled from the systems and the
  heat flux calculations were performed under microcanonical conditions, and
  recorded every 5 fs intervals. For every system, $\kappa$ was estimated by
  averaging over 10-20 independent runs each with simulation times of 20-30 ns,
  and the uncertainty is computed from the standard deviation of the
  independent dataset.}\BibitemShut {Stop}%
\end{thebibliography}

\newpage
\noindent {\bf Supplemental Information}

We report our study on the phonon transport properties in silicon membranes with nanopatterned surfaces with thickness ranging from 1 nm to 20 nm. We highlight the impact of surface resonances on phonon properties and compute the spectral contribution to thermal conductivities. We compute the phonon eigenfrequencies $(\omega_{{\bf q},s})$ and eigenvectors ${\bf e}({\bf q},s)$ within the harmonic approximation by direct diagonalization of the dynamical matrix $(D({\bf q}))$ of a nanostructured membrane super cell with Fourier expansion along the symmetry directions of the two-dimensional Brillouin zone, where ${\bf q}$ is the phonon wavevector and $s$ the polarization. The phonon group velocities are defined as $v_{{\bf q},s} = d\omega_{{\bf q},s}/d{\bf q}$ and computed from the dynamical matrix using first order perturbation theory, $v_{{\bf q},s} = \frac{1}{\omega_{{\bf q},s}} \langle {\bf q},s | \frac{dD({\bf q})}{d{\bf q}}|{\bf q},s \rangle$. 

\begin{figure}
\includegraphics[width=\linewidth]{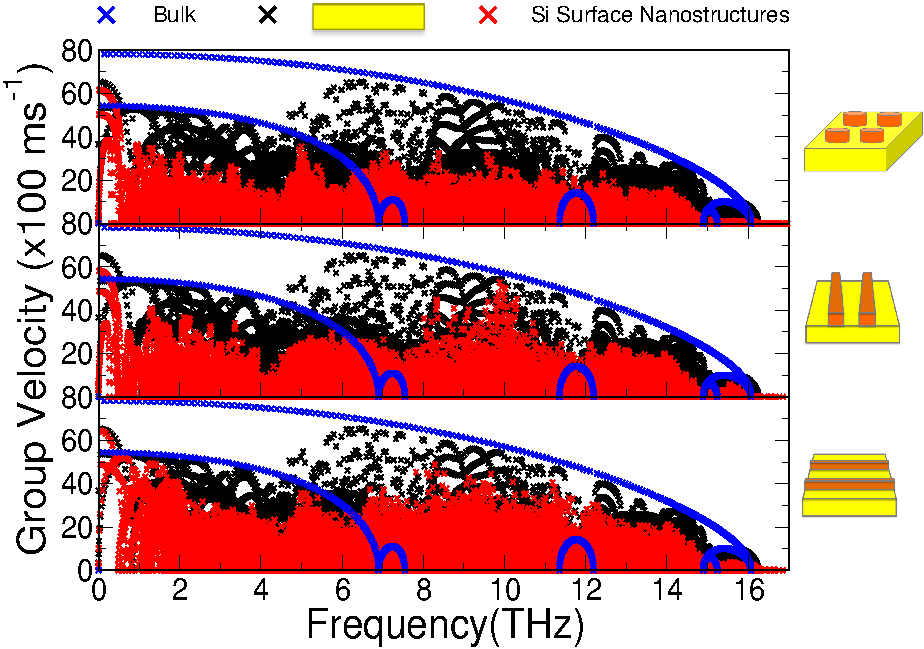}
\caption{Effect of geometry of the nanostructures on the phonon group velocities in nanostructured silicon membranes. The blue x's represent the group velocities in the bulk silicon. The black x's show that the group velocities are reduced due to dimensional reduction in the 3 nm-thick membrane. The orange and green x's represent 3nm-thick silicon membrane with Si nanostructures with geometries depicted by the cartoons in the rightmost column. The surface nanostructures are largely responsible to reduce phonon group velocities in membranes.}
\label{fig:gvel}
\end{figure}

\begin{figure}[b]
\includegraphics[width=\linewidth]{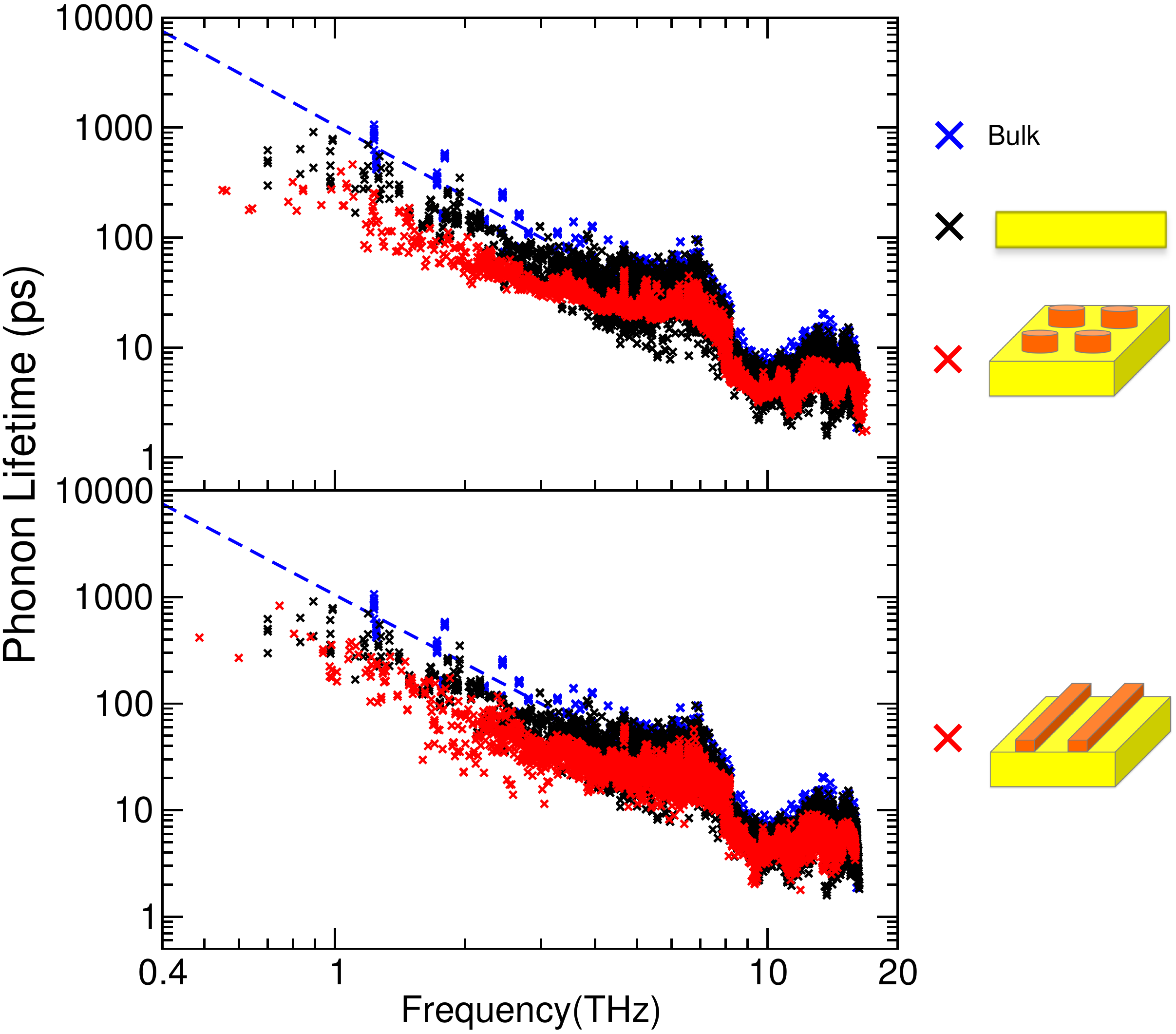}
\caption{Computed relaxation lifetimes in nanostructured silicon membranes: the black and red lines represent membranes with smooth surface and surfaces with silicon nanostructures, respectively. The relaxation times in bulk silicon are shown in blue, for reference. Surface nanostructures reduce the lifetimes of phonon modes in different frequency regions.}
\label{fig:lifetimes_membranes}
\end{figure}

\end{document}